\documentclass[preprint,prl,superscriptaddress,showpacs]{revtex4}
\usepackage{amsmath}

\newcommand{\nc}{\newcommand}
\nc{\rnc}{\renewcommand}
\nc{\ket}[1]{\left | \, #1 \right \rangle}
\nc{\bra}[1]{\left \langle #1 \, \right |}
\nc{\proj}[1]{\ket{#1}\bra{#1}}
\nc{\braket}[2]{\langle\, #1\,|\,#2\,\rangle}
\nc{\hilb}{\mathcal{H}}
\nc{\inprod}[2]{\braket{#1}{#2}}
\def\Id{{\mathbf 1}}
\nc{\half}{\mbox{$\frac{1}{2}$}}
\nc{\third}{\mbox{$\frac{1}{3}$}}
\nc{\sixth}{\mbox{$\frac{1}{6}$}}

\begin{document}
\title{Fidelity of Single Qubit Maps}
\author{Mark D. Bowdrey}
\affiliation{Centre for Quantum Computation, Clarendon Laboratory,
University of Oxford, Parks Road, OX1 3PU, United Kingdom}

\author{Daniel K. L. Oi}
\affiliation{Centre for Quantum Computation, Clarendon Laboratory,
University of Oxford, Parks Road, OX1 3PU, United Kingdom}

\author{Anthony~J.~Short}
\affiliation{Centre for Quantum Computation, Clarendon Laboratory,
University of Oxford, Parks Road, OX1 3PU, United Kingdom}

\author{Konrad Banaszek}
\affiliation{Centre for Quantum Computation, Clarendon Laboratory,
University of Oxford, Parks Road, OX1 3PU, United Kingdom}

\author{Jonathan A. Jones}\email{jonathan.jones@qubit.org}
\thanks{to whom correspondence should be addressed at the
Clarendon Laboratory} \affiliation{Centre for Quantum Computation,
Clarendon Laboratory, University of Oxford, Parks Road, OX1 3PU,
United Kingdom} \affiliation{Oxford Centre for Molecular Sciences,
Central Chemistry Laboratory, University of Oxford, South Parks
Road, OX1 3QH, United Kingdom
\\email: jonathan.jones@qubit.org
\\FAX: +44 1865 272387}
\date{\today}
\begin{abstract}
We describe a simple way of characterizing the average fidelity
between a unitary (or anti-unitary) operator and a general
operation on a single qubit, which only involves calculating the
fidelities for a few pure input states, and discuss possible
applications to experimental techniques including Nuclear Magnetic
Resonance (NMR).
\end{abstract}
\maketitle

In quantum information theory \cite{divincenzo00} it is often
useful to compare the effects of two processes applied to a
quantum system.  The basic building blocks of quantum information
processing are transformations (maps) on two level quantum systems
known as quantum bits or qubits.  Ideally, we would like to be
able to compare any two single qubit maps, but unfortunately this
is not always straightforward.  The comparison is, however, much
simpler if one map is unitary or anti-unitary.  A natural approach
to compare two maps is to calculate the state fidelity of their
output states given identical inputs.  The Uhlmann state fidelity
of two density operators ($\rho_1, \rho_2$) is given by
\cite{uhlmann76} \begin{equation} \label{uhlmannfid}
F(\rho_{1},\rho_{2})=\left(\text{Tr}\left(\sqrt{\sqrt{\rho_{1}}\rho_{2}\sqrt{\rho_{1}}}\,\right)\right)^2.
\end{equation}
This may be interpreted as the maximal overlap of all
purifications of $\rho_1$ and $\rho_2$.  Under a unitary or
anti-unitary transformation, a pure input state maps to a pure
output state and in this case we can simplify the state fidelity
(\ref{uhlmannfid}) to \cite{smolin98}
\begin{equation}
F(\proj{\psi},\rho)=\text{Tr}\left(\proj{\psi}\rho\right).
\end{equation}
The state fidelity of a unitary (or anti-unitary) map $U$ and a
general linear, trace-preserving, transformation $\mathcal{M}$
acting on an initially pure state $\proj{\psi}$ is given by
\begin{equation}
F_{\proj{\psi}}=\text{Tr}\bigl(
U\proj{\psi}U^{\dagger}\mathcal{M}\left[\proj{\psi}\right]\bigr).
\end{equation}
The average map fidelity can then be defined by integrating over all pure input states,
\begin{equation}\label{eqDefFid}
\bar{F}=\frac{1}{4\pi} \int\! F_{\proj{\psi}} \,\text{d}\Omega,
\end{equation}
(where the integral is over the surface of the Bloch sphere) and
this definition is widely used
\cite{hardy01,song01,fiurasek01,moor01}.  There is, however, a
simplification: using the fact that $\proj{\psi}$ can be written
in terms of the Pauli spin matrices and the identity matrix
\cite{zoller97},
\begin{equation}
\begin{split}
\proj{\psi_{\theta,\phi}} & = \half\left(\Id + \left(
\begin{array}{cc}
\cos\theta & \sin\theta e^{-i\phi} \\
 \sin\theta e^{i\phi} & -\cos\theta
\end{array}
\right)
\right)
\\
&=\half\,(\sigma_{0}+\sin\theta\,\cos\phi\,\sigma_{x}+\sin\theta\,\sin\phi\,\sigma_{y}+\cos\theta\,\sigma_{z})\\
&= \!\!\!\!\!\sum_{j=0,x,y,z}\!\!\!\!\! c_{j}(\theta,\phi)
\frac{\sigma_{j}}{2}
\end{split}
\end{equation}
we can now express equation (\ref{eqDefFid}) as,
\begin{equation}
\begin{split}
\bar{F} & = \frac{1}{4\pi}\int_{\theta=0}^{\pi}
\int_{\phi=0}^{2\pi} \text{Tr}\left( U \left[ \sum_{j}c_{j}(\theta,\phi)
\frac{\sigma_{j}}{2}\right] U^{\dagger}\,
\mathcal{M}\!\left[\sum_{k}c_{k}(\theta,\phi) \frac{\sigma_{k}}{2}\right]\right)
\sin{\theta}\, \text{d}\phi\, \text{d}\theta\\
        & =
\sum_{jk} \left(
  \frac{1}{4\pi}\int_{\theta} \int_{\phi} c_{j}c_{k}\sin\theta\, \text{d}\phi\, \text{d}\theta
\right) \text{Tr}\left( U\frac{\sigma_{j}}{2} U^{\dagger} \,
\mathcal{M} \!\left[ \frac{\sigma_{k}}{2} \right] \right)
\end{split}
\end{equation}
where we have used the linearity of $U$ and $\mathcal{M}$. When
integrated over the Bloch sphere the coefficients of the
off-diagonal terms go to zero, while the diagonal terms survive
\cite{bowdrey01}, leaving
\begin{equation}
\label{pauliFid}
\begin{split}
\bar{F} & =
\sum_{jk}\left(\frac{2\delta_{j0}\delta_{k0}+\delta_{jk}}{3}\right)
\text{Tr}\left( U\frac{\sigma_{j}}{2} U^{\dagger} \, \mathcal{M} \!\left[ \frac{\sigma_{k}}{2} \right] \right)\\
        & =
\text{Tr}\left(U\frac{\sigma_{0}}{2}U^{\dagger}\,
\mathcal{M}\!\left[\frac{\sigma_{0}}{2}\right]\right)+
\third\!\!\!\!\sum_{j=x,y,z}\!\!\!\!\text{Tr}\left(U\frac{\sigma_{j}}{2}U^{\dagger}\, \mathcal{M}\!\left[\frac{\sigma_{j}}{2}\right]\right)\\
        & =
\half+
\third\!\!\!\!\sum_{j=x,y,z}\!\!\!\!\text{Tr}\left(U\frac{\sigma_{j}}{2}U^{\dagger}\,
\mathcal{M}\!\left[\frac{\sigma_{j}}{2}\right]\right),
\end{split}
\end{equation}
where we have used the unit trace of $\sigma_0$ and the fact that
$\mathcal{M}$ is trace-preserving.

Expressing the average fidelity in this form may not seem helpful
as the Pauli spin matrices do not represent proper states.
However, in NMR experiments where the states are highly mixed,
single qubit states can be represented by $\{ \half \sigma_x,\half
\sigma_y,\half \sigma_z \}$ \cite{ernst, cory97, jones01} and
therefore we can use equation (\ref{pauliFid}) directly.  One
application of this approach is to characterise the behaviour of
composite rotation sequences \cite{freeman97, levitt86}, which are
widely used in NMR to reduce the effects of systematic errors. In
conventional NMR experiments \cite{freeman97} composite rotations
are used to effect particular motions on the Bloch sphere (such as
inversion, which takes a spin from $+z$ to $-z$), and it suffices
to determine the point-to-point fidelity, but when used in NMR
implementations of quantum computation \cite{cummins00} the
initial state is unknown. One approach used to date is Levitt's
quaternion fidelity \cite{levitt86, cummins00} but this has the
major disadvantage that it can only be used to asses the
theoretical behaviour of a rotation sequence and cannot be
determined by experiment.  The average fidelity approach outlined
above provides a simple approach which can be used for \emph{both}
theoretical \emph{and} experimental studies.

For experimental and theoretical work with pure state techniques
we require a more appropriate form and so we use the substitutions
\begin{equation}
\begin{split}
\frac{\sigma_{j}}{2} & = \frac{\Id+\sigma_{j}}{2}-\frac{\Id}{2}
                       = \rho_{j}-\rho_{0}\\
                     & = \frac{\Id}{2} - \frac{\Id-\sigma_{j}}{2}
                       = \rho_{0} - \rho_{-j},
\end{split}
\end{equation}
where $\rho_{\pm j}$ represents a pure state in the $\pm
j$-direction and $\rho_{0}$ is the maximally mixed state.  This gives the
two equivalent expressions
\begin{equation}\label{eqFidAv}
\bar{F} = \half+\third\!\!\!\!\!\sum_{j=x,y,z}\!\!\!\!\!\left(
\text{Tr}\left(U\rho_{j}U^{\dagger}\,
\mathcal{M}\!\left[\rho_{j}\right]\right)-
\text{Tr}\left(U\rho_{j}U^{\dagger}\,
\mathcal{M}\!\left[\rho_{0}\right]\right) \right)
\end{equation}
\begin{equation}\label{eqFidAv2}
\bar{F} = \half+\third\!\!\!\!\!\sum_{j=x,y,z}\!\!\!\!\!\left(
\text{Tr}\left(U\rho_{-j}U^{\dagger}\,
\mathcal{M}\!\left[\rho_{-j}\right]\right)-
\text{Tr}\left(U\rho_{-j}U^{\dagger}\,
\mathcal{M}\!\left[\rho_{0}\right]\right) \right),
\end{equation}
and taking the average of (\ref{eqFidAv}) and (\ref{eqFidAv2})
yields,
\begin{equation}\begin{split}
\bar{F} & = \half+\sixth\!\!\!\!\sum_{j=x,y,z}\!\!\!\!\left(
\text{Tr}\left(U\rho_{j}U^{\dagger}\,
\mathcal{M}\!\left[\rho_{j}\right]\right)+
\text{Tr}\left(U\rho_{-j}U^{\dagger}\,
\mathcal{M}\!\left[\rho_{-j}\right]\right)-
\text{Tr}\left(U(\rho_{j}+\rho_{-j})U^{\dagger}\,
\mathcal{M}\!\left[\rho_{0}\right]\right)
\right)\\
        & =
\half+\sixth\!\!\!\!\sum_{j=x,y,z}\!\!\!\!\left(
\text{Tr}\left(U\rho_{j}U^{\dagger}\,
\mathcal{M}\!\left[\rho_{j}\right]\right)+
\text{Tr}\left(U\rho_{-j}U^{\dagger}\,
\mathcal{M}\!\left[\rho_{-j}\right]\right)- 2\,
\text{Tr}\left(U\rho_{0}U^{\dagger}\,
\mathcal{M}\!\left[\rho_{0}\right]\right)
\right)\\
        & =
\half+\sixth\!\!\!\!\sum_{j=x,y,z}\!\!\!\!\left(
\text{Tr}\left(U\rho_{j}U^{\dagger}\,
\mathcal{M}\!\left[\rho_{j}\right]\right)+
\text{Tr}\left(U\rho_{-j}U^{\dagger}\,
\mathcal{M}\!\left[\rho_{-j}\right]\right)-1
\right)\\
        & =
\sixth\!\!\!\!\sum_{j=\pm x,\pm y,\pm z}\!\!\!\!\left(
\text{Tr}\left(U\rho_{j}U^{\dagger}\,
\mathcal{M}\!\left[\rho_{j}\right]\right) \right).
\end{split}\end{equation}
Hence, the fidelity of the map $\mathcal{M}$ with the unitary or
anti-unitary map $U$ can be calculated by simply averaging the
fidelities of the six axial pure states on the Bloch sphere,
$\{\rho_{+x},\rho_{-x},\rho_{+y},\rho_{-y},\rho_{+z},\rho_{-z}\}$.
We note that the average map fidelity ($\bar{F}$) can in fact be
characterized by only four pure states,
$\{\frac{1}{2}(1+\frac{1}{\sqrt{3}}(+\sigma_x + \sigma_y +
\sigma_z)), \frac{1}{2}(1+\frac{1}{\sqrt{3}}(-\sigma_x - \sigma_y
+ \sigma_z)), \frac{1}{2}(1+\frac{1}{\sqrt{3}}(-\sigma_x +
\sigma_y - \sigma_z)), \frac{1}{2}(1+\frac{1}{\sqrt{3}}(+\sigma_x
- \sigma_y - \sigma_z))\}$.  Indeed, the fidelity can be
characterized using any four pure states forming a regular
tetrahedron, or any six forming a regular octahedron; however the
pure states at the six cardinal points provide a particularly
natural approach.

An obvious application of this result is to compare a desired
unitary operation with its actual implementation that (due to
experimental imperfections) may be more closely represented by a
superoperator.  A practical advantage of characterizing the
fidelity by just testing six states is that this approach provides
a simple means to verify the map fidelity by experiment.
Similarly, we can also use this result to calculate the fidelity
of a unitary or superoperator approximation to an anti-unitary map
\cite{hardy01} in a convenient and intuitive manner.

\begin{acknowledgments}
We thank E.~Galv\~{a}o and L.~Hardy for helpful conversations.
M.D.B. and A.J.S thank EPSRC (UK) for financial support.  D.K.L.O.
thanks CESG (UK) and QAIP (contract IST-1999-11234) for financial
support.  K.B. thanks EQUIP (contract IST-1999-11053) for
financial support. J.A.J. is a Royal Society University Research
Fellow. This is in part a contribution from the Oxford Centre for
Molecular Sciences, which is supported by the UK EPSRC, BBSRC, and
MRC.
\end{acknowledgments}

\end{document}